\documentclass[runningheads]{llncs}
\usepackage{graphicx}
\usepackage{amsfonts}
\usepackage{amssymb}
\usepackage{hyperref}
\usepackage{xcolor}
\usepackage{tikz}
\usepackage{tikz-3dplot}
\usepackage{marvosym}

\newcommand{\rcosmo}{{\bf rcosmo}\,}

\usepackage{floatrow} 
\hypersetup{ 
    colorlinks=true, 
    linkcolor=blue,
    filecolor=magenta,      
    urlcolor=blue,
    citecolor=blue
}

\newcommand{\R}{\mathbb{R}}

\newcommand{\set}[1]{{\left\{#1\right\}}}

\newcommand{\deq}{\mathbin{\raise0.4pt\hbox{:}{=}}}

\begin{document}
\title{Spherical data handling and analysis with R~package rcosmo\thanks{This research was partially supported under the Australian
Research Council's Discovery Project DP160101366.}}

%
%
\author{Daniel Fryer\orcidID{0000-0001-6032-0522} \and
Andriy Olenko\raisebox{1mm}{\Letter\ }\orcidID{0000-0002-0917-7000}}
\authorrunning{D.Fryer and A.Olenko}
%
\institute{Department of Mathematics and Statistics,
  La Trobe University, VIC, 3086
\email{d.fryer@latrobe.edu.au}\\
\email{a.olenko@latrobe.edu.au}}
\maketitle              
\begin{abstract}
The R package \rcosmo was developed for handling and ana\-lysing Hierarchical Equal Area isoLatitude Pixelation(HEALPix) and Cosmic Microwave Background(CMB) radiation data. It has more than 100  functions. \rcosmo  was initially developed for CMB, but also can be used for other spherical data. This paper discusses transformations into  \rcosmo formats and handling of three types of non-CMB data: continuous geographic, point pattern and star-shaped.  For each type of data we provide a brief description of the corresponding statistical model, data example and ready-to-use R code. Some statistical functionality of \rcosmo is demonstrated for the example data converted into   the HEALPix format. The paper can serve as the first practical guideline to transforming data into the HEALPix format and statistical analysis with \rcosmo for geo-statisticians, GIS and R users and  researches dealing with spherical data in non-HEALPix formats.
\keywords{spatial statistics  \and HEALPix \and random field \and spatial point process \and directional \and star-shaped}
\end{abstract}
\section{Introduction}
The package \rcosmo was developed to offer the functionality needed to handle and analyse Hierarchical Equal Area isoLatitude Pixelation (HEALPix) and Cosmic Microwave Background (CMB) radiation data. Comprehensive  software packages for working with HEALPix data are available in Python and MATLAB, see,  for example, \cite{healpix},  \cite{healpix1} and \cite{healpix2}. The main aim of \rcosmo was to provide a convenient access to big CMB data and HEALPix functionality for R users. 

The package has more than 100  functions. They can be broadly divided into four groups: 
\begin{itemize}
\item CMB and HEALPix data holding, subsetting and visualization,
\item HEALPix structure operations,
\item geometric methods, 
\item statistical methods.
\end{itemize}
The detailed summary of \rcosmo structure, HEALPix, geometric and statistical functionality is provided in \cite{Fryer 2}. Technical description and examples of the core \rcosmo functions can be found in the package documentation on CRAN~\cite{Fryer}. This paper addresses a different important problem of using \rcosmo for spherical-type data in non-HEALPix formats.

\rcosmo  was initially developed for CMB data that are HEALPix indexed and can be represented as GIS (geographic information system) raster images and modelled as random fields. However, there are other spherical coordinate systems and statistical models. Spherical data are of main interest for geosciences, environmetric and  biological studies, but most of researches in these fields are not aware about advantages of the HEALPix data structure or do not have ready-to-use R code to transform their data into HEALPix formats.  The aim of this publication is to demonstrate how to deal with three different types of non-CMB/non-HEALPix data.  First, we consider continuous geo-referenced observations that are modelled by random fields. The second type of data are discrete spherical data that can be given as realisations of spatial point processes. Finally, the analysis of irregularly star-shaped geometric bodies is presented. Real data examples and illustrations of basic \rcosmo statistical functions are given for each of the three types.

To reproduce the results of this paper the current version of the package \rcosmo  can be installed from CRAN. A development release is available from GitHub (\url{https://github.com/frycast/rcosmo}). A reproducible version of the code and the data used in this paper  are available in the folder "Research materials" from the  website~\url{https://sites.google.com/site/olenkoandriy/}.

\section{Coordinate systems for spherical data representation}
The HEALPix format has numerous advantages, compared to other spherical data representations: equal area pixels,  hierarchical tessellations of the sphere and iso-latitude rings of pixels. It is used for an efficient organization of spherical data in a computer memory and providing fast spherical harmonic transforms, search and numerical analysis of spherical data. 

The HEALPix representation is the key element for indexing spherical data in \rcosmo. This section recalls the main spherical coordinate systems and introduces basics of HEALPix. The following sections will demonstrate conversion from  different data representations into the HEALPix format.

To index locations of observations the vast majority of spatial applications dealing with spherical data use one of the following coordinate system: Cartesian, geographic, spherical or HEALPix. The Cartesian and spherical coordinate systems often appear in mathematical description of models. The geographic coordinates are the main indexing tools in GIS, Earth and planetary sciences, while HEALPix has become very popular  in recent cosmological research dealing with CMB data. 

For simplicity, this section considers the unit sphere with radius 1. Using the {\it Cartesian coordinate system} a location on the sphere is specified by a triplet $(x,y,z),$ where   $x,y,z \in \R$ and $||(x,y,z)|| = \sqrt{x^2+y^2+z^2} = 1$. The {\it spherical coordinates} $(\theta, \varphi)$ of a point are obtained from $(x,y,z)$ by inverting the three equations
\[
x = \sin(\theta)\cos(\varphi), \;\; y =  \sin(\theta)\sin(\varphi), \;\; z = \cos(\theta),
\]
where $\theta \in [0,\pi]$ and $\varphi \in [0,2\pi).$


For a point with the spherical coordinates $(\theta, \varphi)$ its geographic coordinates will be written as $(\theta_G, \varphi_G).$ {\it Geographic coordinates} are obtained from spherical coordinates by setting 
\[
\varphi_G = \left\{\begin{array}{ll}
       \varphi, & \mbox{ for } \varphi \in [0,\pi], \\
        \varphi - 2\pi, & \mbox{ for } \varphi \in (\pi, 2\pi),
        \end{array}\right. \quad \mbox{and} \quad \theta_G = \frac{\pi}{2} - \theta.
  \]
Thus $\varphi_G \in (-\pi, \pi]$ and $\theta_G \in [-\pi/2, \pi/2]$. When representing Earth's surface in any of the above coordinate systems we align the $x$-axis with the Earth's Prime Meridian, and have the $z$-axis pointing north. Commonly $\varphi_G$ is referred to as longitude and $\theta_G$ is referred to as latitude and the both are often measured in degrees instead of radians. 

HEALPix is a {\it Hierarchical Equal Area Isolatitude Pixelation} of the sphere. Detailed derivations of the HEALPix coordinates can be found in \cite{Gorski}. First, the unit sphere is divided into 12 equatorial base pixels.  A planar projection of the base pixels is given in Figure \ref{healpix-base}. The base pixelation divides the sphere into one equatorial and two polar regions. Referring to the indices shown in Figure \ref{healpix-base}, pixels $5, 6,7$ and $8$ are ``equatorial"; pixels $1,2,3$ and $4$ are ``north polar"; and pixels $9,10,11$ and $12$ are ``south polar." 

\pgfmathsetmacro{\sqA}{0}
\pgfmathsetmacro{\sqB}{2}
\pgfmathsetmacro{\sqC}{4}
\pgfmathsetmacro{\sqD}{6}

\pgfmathsetmacro{\indA}{2}
\pgfmathsetmacro{\indB}{4}
\pgfmathsetmacro{\indC}{6}
\pgfmathsetmacro{\indD}{8}
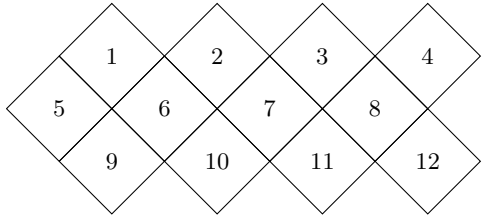
\begin{figure}[htb]
\begin{center}
\begin{tikzpicture}[scale = 0.7]
\draw (\sqA,0) -- (\sqA+1,1) -- (\sqA+2,0) -- (\sqA+1,-1) -- (\sqA,0);
\draw (\sqB,0) -- (\sqB+1,1)  -- (\sqB+2,0) -- (\sqB+1,-1) -- (\sqB,0);
\draw (\sqC,0) -- (\sqC+1,1) -- (\sqC+2,0) -- (\sqC+1,-1) -- (\sqC,0);
\draw (\sqD,0) -- (\sqD+1,1) -- (\sqD+2,0) -- (\sqD+1,-1) -- (\sqD,0);

\draw (\sqA+1,0+1) -- (\sqA+1+1,1+1) -- (\sqA+2+1,0+1) -- (\sqA+1+1,-1+1) -- (\sqA+1,0+1);
\draw (\sqB+1,0+1) -- (\sqB+1+1,1+1) -- (\sqB+2+1,0+1) -- (\sqB+1+1,-1+1) -- (\sqB+1,0+1);
\draw (\sqC+1,0+1) -- (\sqC+1+1,1+1) -- (\sqC+2+1,0+1) -- (\sqC+1+1,-1+1) -- (\sqC+1,0+1);
\draw (\sqD+1,0+1) -- (\sqD+1+1,1+1) -- (\sqD+2+1,0+1) -- (\sqD+1+1,-1+1) -- (\sqD+1,0+1);

\draw (\sqA+1,0-1) -- (\sqA+1+1,1-1) -- (\sqA+2+1,0-1) -- (\sqA+1+1,-1-1) -- (\sqA+1,0-1);
\draw (\sqB+1,0-1) -- (\sqB+1+1,1-1) -- (\sqB+2+1,0-1) --  (\sqB+1+1,-1-1) -- (\sqB+1,0-1);
\draw (\sqC+1,0-1) -- (\sqC+1+1,1-1) -- (\sqC+2+1,0-1) -- (\sqC+1+1,-1-1) -- (\sqC+1,0-1);
\draw (\sqD+1,0-1) -- (\sqD+1+1,1-1) -- (\sqD+2+1,0-1) -- (\sqD+1+1,-1-1) -- (\sqD+1,0-1);

\draw (\indA-1, 0)  node {5};
\draw (\indA, -1)    node {9};
\draw (\indA, +1)   node {1};

\draw (\indB-1, 0)  node {6};
\draw (\indB, -1)    node {10};
\draw (\indB, +1)   node {2};

\draw (\indC-1, 0)  node {7};
\draw (\indC, -1)    node {11};
\draw (\indC, +1)   node {3};

\draw (\indD-1, 0)  node {8};
\draw (\indD, -1)    node {12};
\draw (\indD, +1)   node {4};
\end{tikzpicture}
\caption{HEALPix base pixel planar projection as 12 squares.} \label{healpix-base}
\end{center} 
\end{figure}
The base pixelation is defined to have the resolution parameter $j = 0$. For resolution $j = 1,$ each base pixel is subdivided into 4 equiareal child pixels. This process is repeated for higher resolutions with each pixel at resolution $j = k$ being one of 4 child pixels from the subdivision of its parent pixel in resolution $j = k - 1$. At any resolution $j$, the number $N_{{s}}$ of pixels per base pixel edge is $N_{{s}} = 2^j$ and the total number of pixels is $T = 12N_{{s}}^2$.  

During this subdivision, pixel boundary and centre locations are chosen in such a way that all pixel centres lie on $4N_{{s}} - 1$ rings of constant latitude, making it easy to implement various mathematical methods, in particular  Fourier transforms with spherical harmonics. Pixel indices are then assigned to child pixels in one of two ways, known as the ``ring" and ``nested" \textit{ordering schemes}. In the ring ordering scheme, indices are assigned in the increasing order from east to west along isolatitude rings, and then increasing north to south, as in the example shown in Figure~\ref{healpix-ring}. In the nested ordering scheme  the children of base pixel $b \in \set{1,2,\ldots,12}$ are labelled with $T/12$ consecutive labels as shown in Figure~\ref{healpix-nested}. This nested ordering scheme allows efficient implementation of local operations such as nearest-neighbour searches.

\pgfmathsetmacro{\sqA}{0}
\pgfmathsetmacro{\sqB}{2}
\pgfmathsetmacro{\sqC}{4}
\pgfmathsetmacro{\sqD}{6}

\pgfmathsetmacro{\indA}{2}
\pgfmathsetmacro{\indB}{4}
\pgfmathsetmacro{\indC}{6}
\pgfmathsetmacro{\indD}{8}
\begin{figure}[htb]\vspace{-4mm}
\begin{center}
\begin{tikzpicture}[scale = 1]
\draw (\sqA,0) -- (\sqA+1,1) -- (\sqA+2,0) -- (\sqA+1,-1) -- (\sqA,0);
\draw (\sqB,0) -- (\sqB+1,1)  -- (\sqB+2,0) -- (\sqB+1,-1) -- (\sqB,0);
\draw (\sqC,0) -- (\sqC+1,1) -- (\sqC+2,0) -- (\sqC+1,-1) -- (\sqC,0);
\draw (\sqD,0) -- (\sqD+1,1) -- (\sqD+2,0) -- (\sqD+1,-1) -- (\sqD,0);

\draw (\sqA+1,0+1) -- (\sqA+1+1,1+1) -- (\sqA+2+1,0+1) -- (\sqA+1+1,-1+1) -- (\sqA+1,0+1);
\draw (\sqB+1,0+1) -- (\sqB+1+1,1+1) -- (\sqB+2+1,0+1) -- (\sqB+1+1,-1+1) -- (\sqB+1,0+1);
\draw (\sqC+1,0+1) -- (\sqC+1+1,1+1) -- (\sqC+2+1,0+1) -- (\sqC+1+1,-1+1) -- (\sqC+1,0+1);
\draw (\sqD+1,0+1) -- (\sqD+1+1,1+1) -- (\sqD+2+1,0+1) -- (\sqD+1+1,-1+1) -- (\sqD+1,0+1);

\draw (\sqA+1,0-1) -- (\sqA+1+1,1-1) -- (\sqA+2+1,0-1) -- (\sqA+1+1,-1-1) -- (\sqA+1,0-1);
\draw (\sqB+1,0-1) -- (\sqB+1+1,1-1) -- (\sqB+2+1,0-1) --  (\sqB+1+1,-1-1) -- (\sqB+1,0-1);
\draw (\sqC+1,0-1) -- (\sqC+1+1,1-1) -- (\sqC+2+1,0-1) -- (\sqC+1+1,-1-1) -- (\sqC+1,0-1);
\draw (\sqD+1,0-1) -- (\sqD+1+1,1-1) -- (\sqD+2+1,0-1) -- (\sqD+1+1,-1-1) -- (\sqD+1,0-1);

\draw (\indA, +1+0.5)   node {1};
\draw (\indA-0.5, +1)   node {5};
\draw (\indA+0.5, +1)   node {6};
\draw (\indA, +1-0.5)   node {14};

\draw (\indB        , +1+0.5)   node {2};
\draw (\indB-0.5 , +1)           node {7};
\draw (\indB+0.5, +1)          node {8};
\draw (\indB        , +1-0.5)    node {16};

\draw (\indC        , +1+0.5)   node {3};
\draw (\indC-0.5 , +1)           node {9};
\draw (\indC+0.5, +1)          node {10};
\draw (\indC        , +1-0.5)    node {18};

\draw (\indD        , +1+0.5)   node {4};
\draw (\indD-0.5 , +1)           node {11};
\draw (\indD+0.5, +1)          node {12};
\draw (\indD        , +1-0.5)    node {20};

\draw (\indA-1, 0.5)  node {13};
\draw (\indA-1-0.5,0)  node {21};
\draw (\indA-1+0.5,0)  node {22};
\draw (\indA-1, -0.5)  node {29};

\draw (\indB-1, 0.5)  node {15};
\draw (\indB-1-0.5,0)  node {23};
\draw (\indB-1+0.5,0)  node {24};
\draw (\indB-1, -0.5)  node {31};

\draw (\indC-1, 0.5)  node {17};
\draw (\indC-1-0.5,0)  node {25};
\draw (\indC-1+0.5,0)  node {26};
\draw (\indC-1, -0.5)  node {33};

\draw (\indD-1, 0.5)  node {19};
\draw (\indD-1-0.5,0)  node {27};
\draw (\indD-1+0.5,0)  node {28};
\draw (\indD-1, -0.5)  node {35};

\draw (\indA, -1+0.5)    node {30};
\draw (\indA-0.5, -1)     node {37};
\draw (\indA+0.5, -1)    node {38};
\draw (\indA, -1-0.5)     node {45};

\draw (\indB, -1+0.5)    node {32};
\draw (\indB-0.5, -1)     node {39};
\draw (\indB+0.5, -1)    node {40};
\draw (\indB, -1-0.5)     node {46};

\draw (\indC, -1+0.5)    node {34};
\draw (\indC-0.5, -1)     node {41};
\draw (\indC+0.5, -1)    node {42};
\draw (\indC, -1-0.5)     node {47};

\draw (\indD, -1+0.5)    node {36};
\draw (\indD-0.5, -1)     node {43};
\draw (\indD+0.5, -1)    node {44};
\draw (\indD, -1-0.5)     node {48};

\draw (0.5, -0.5) -- (0.5 + 2, -0.5 + 2);
\draw (1.5, -1.5) -- (1.5+3, -1.5+3);
\draw (3.5, -1.5) --  (3.5+3, -1.5+3);
\draw (5.5, -1.5) --  (5.5+3, -1.5+3);
\draw (7.5, -1.5) --  (7.5+1, -1.5+1);

\draw (0.5, 0.5) --  (0.5+2, 0.5-2);
\draw (1.5, 1.5) --  (1.5+3, 1.5-3);
\draw (3.5, 1.5) --  (3.5+3, 1.5-3);
\draw (5.5, 1.5) --  (5.5+3, 1.5-3);
\draw (7.5, 1.5) --  (7.5+1, 1.5-1);
\end{tikzpicture}
\caption{HEALPix pixelation at resolution $j = 1$ in ring ordering scheme.} \label{healpix-ring}
\end{center} 
\end{figure}
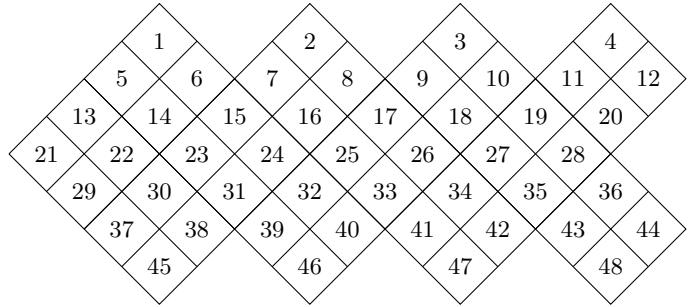

\pgfmathsetmacro{\sqA}{0}
\pgfmathsetmacro{\sqB}{2}
\pgfmathsetmacro{\sqC}{4}
\pgfmathsetmacro{\sqD}{6}
 
\pgfmathsetmacro{\indA}{2}
\pgfmathsetmacro{\indB}{4}
\pgfmathsetmacro{\indC}{6}
\pgfmathsetmacro{\indD}{8}
\begin{figure}[htb]\vspace{-6mm}
\begin{center}
\begin{tikzpicture}[scale = 1]
\draw (\sqA,0) -- (\sqA+1,1) -- (\sqA+2,0) -- (\sqA+1,-1) -- (\sqA,0);
\draw (\sqB,0) -- (\sqB+1,1) -- (\sqB+2,0) -- (\sqB+1,-1) -- (\sqB,0);
\draw (\sqC,0) -- (\sqC+1,1) -- (\sqC+2,0) -- (\sqC+1,-1) -- (\sqC,0);
\draw (\sqD,0) -- (\sqD+1,1) -- (\sqD+2,0) -- (\sqD+1,-1) -- (\sqD,0);
 
\draw (\sqA+1,0+1) -- (\sqA+1+1,1+1) -- (\sqA+2+1,0+1) -- (\sqA+1+1,-1+1) -- (\sqA+1,0+1);
\draw (\sqB+1,0+1) -- (\sqB+1+1,1+1) -- (\sqB+2+1,0+1) -- (\sqB+1+1,-1+1) -- (\sqB+1,0+1);
\draw (\sqC+1,0+1) -- (\sqC+1+1,1+1) -- (\sqC+2+1,0+1) -- (\sqC+1+1,-1+1) -- (\sqC+1,0+1);
\draw (\sqD+1,0+1) -- (\sqD+1+1,1+1) -- (\sqD+2+1,0+1) -- (\sqD+1+1,-1+1) -- (\sqD+1,0+1);
 
\draw (\sqA+1,0-1) -- (\sqA+1+1,1-1) -- (\sqA+2+1,0-1) -- (\sqA+1+1,-1-1) -- (\sqA+1,0-1);
\draw (\sqB+1,0-1) -- (\sqB+1+1,1-1) -- (\sqB+2+1,0-1) -- (\sqB+1+1,-1-1) -- (\sqB+1,0-1);
\draw (\sqC+1,0-1) -- (\sqC+1+1,1-1) -- (\sqC+2+1,0-1) -- (\sqC+1+1,-1-1) -- (\sqC+1,0-1);
\draw (\sqD+1,0-1) -- (\sqD+1+1,1-1) -- (\sqD+2+1,0-1) -- (\sqD+1+1,-1-1) -- (\sqD+1,0-1);
 
\draw (\indA-1, 0.5) node {20};
\draw (\indA-1-0.5,0) node {19};
\draw (\indA-1+0.5,0) node {18};
\draw (\indA-1, -0.5) node {17};
 
\draw (\indA, -1+0.5) node {36};
\draw (\indA-0.5, -1) node {35};
\draw (\indA+0.5, -1) node {34};
\draw (\indA, -1-0.5) node {33};
 
\draw (\indA, +1+0.5) node {4};
\draw (\indA-0.5, +1) node {3};
\draw (\indA+0.5, +1) node {2};
\draw (\indA, +1-0.5) node {1};
 
\draw (\indB-1, 0.5) node {24};
\draw (\indB-1-0.5,0) node {23};
\draw (\indB-1+0.5,0) node {22};
\draw (\indB-1, -0.5) node {21};
 
\draw (\indB, -1+0.5) node {40};
\draw (\indB-0.5, -1) node {39};
\draw (\indB+0.5, -1) node {38};
\draw (\indB, -1-0.5) node {37};
 
\draw (\indB , +1+0.5) node {8};
\draw (\indB-0.5 , +1) node {7};
\draw (\indB+0.5, +1) node {6};
\draw (\indB , +1-0.5) node {5};
 
\draw (\indC-1, 0.5) node {28};
\draw (\indC-1-0.5,0) node {27};
\draw (\indC-1+0.5,0) node {26};
\draw (\indC-1, -0.5) node {25};
 
\draw (\indC, -1+0.5) node {44};
\draw (\indC-0.5, -1) node {43};
\draw (\indC+0.5, -1) node {42};
\draw (\indC, -1-0.5) node {41};
 
\draw (\indC , +1+0.5) node {12};
\draw (\indC-0.5 , +1) node {11};
\draw (\indC+0.5, +1) node {10};
\draw (\indC , +1-0.5) node {9};
 
\draw (\indD-1, 0.5) node {32};
\draw (\indD-1-0.5,0) node {31};
\draw (\indD-1+0.5,0) node {30};
\draw (\indD-1, -0.5) node {29};
 
\draw (\indD, -1+0.5) node {48};
\draw (\indD-0.5, -1) node {47};
\draw (\indD+0.5, -1) node {46};
\draw (\indD, -1-0.5) node {45};
 
\draw (\indD , +1+0.5) node {16};
\draw (\indD-0.5 , +1) node {15};
\draw (\indD+0.5, +1) node {14};
\draw (\indD , +1-0.5) node {13};
 
\draw (0.5, -0.5) -- (0.5 + 2, -0.5 + 2);
\draw (1.5, -1.5) -- (1.5+3, -1.5+3);
\draw (3.5, -1.5) -- (3.5+3, -1.5+3);
\draw (5.5, -1.5) -- (5.5+3, -1.5+3);
\draw (7.5, -1.5) -- (7.5+1, -1.5+1);
 
\draw (0.5, 0.5) -- (0.5+2, 0.5-2);
\draw (1.5, 1.5) -- (1.5+3, 1.5-3);
\draw (3.5, 1.5) -- (3.5+3, 1.5-3);
\draw (5.5, 1.5) -- (5.5+3, 1.5-3);
\draw (7.5, 1.5) -- (7.5+1, 1.5-1);
\end{tikzpicture}
\caption{HEALPix pixelation at resolution $j = 1$ in nested ordering scheme.} \label{healpix-nested}
\end{center} 
\end{figure}
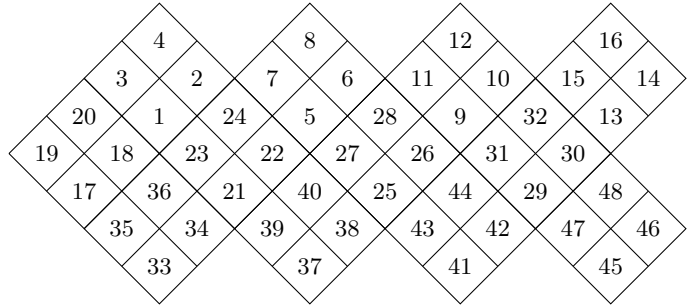

\section{Continuous geographic data}\label{sec3}
In this section we demonstrate how \rcosmo can be applied to handle continuous geo-references observations. Such observations are usually collected over dense geographic grids or obtained as results of spatial interpolation or smoothing.  Continuous geographic data are common in meteorology, for example, maps with temperature, precipitation, wind direction, or atmospheric pressure values.  Other examples of continuous data are land elevations, heights above mean sea level, and ground-level ozone measurements. There are also other numerous environmetrics examples.  These data are usually represented and visualised as topographic/contour maps or GIS raster images. In geographic applications it is often assumed that the Earth has a spherical shape with a radius about 6378~km, but with an elevation that departs from  this sphere in a very irregular manner. Therefore, most applied methods for the above data are based on spherical statistical models.

Traditionally theoretical models that are used for the continuous type of data are called {\it random fields} in statistics or {\it spatially dependent variables} in geostatistics. Below we introduce basic notations and background by reviewing some results about spherical random fields, see more details in \cite{MarinucciPeccati11} and  \cite{Yadrenko}.

We will denote a 3d sphere with radius 1 by
\[
\mathbb{S}^2=\left\{ \mathbf{x}\in \mathbb{R}^{3}:\Vert \mathbf{x}\Vert =1\right\}. 
\]

A spherical random field on a probability space $(\Omega ,\mathcal{F%
},\mathbf{P})$, denoted by%
\[
T=\left\{ T(\theta ,\varphi )=T_{\omega }(\theta ,\varphi ):0\leq \theta
\leq \pi ,\quad 0\leq \varphi \leq 2\pi ,\ \omega \in \Omega \right\} , 
\]%
or $\widetilde{T}=\{\widetilde{T}(\mathbf{x})$ $,$ $\mathbf{x}\in \mathbb{S}^2\},$ is a stochastic function
defined on the sphere $\mathbb{S}^2.$

The field $\widetilde{T}(\mathbf{x})$ is called isotropic (in
the weak sense) on the sphere $\mathbb{S}^2$ if $\mathrm{E}\widetilde{T}%
(\mathbf{x})^{2}<\infty$ and its first and second-order moments are invariant with
respect to the group $SO(3)$ of rotations in $\mathbb{R}^{3},$ i.e. 
\[
\mathbf{E}\widetilde{T}(\mathbf{x})= \mathbf{E}\widetilde{T}(g\mathbf{x}),\quad  \mathbf{E}%
\widetilde{T}(\mathbf{x})\widetilde{T}(\mathbf{y})= \mathbf{E}\widetilde{T}(g\mathbf{x})\widetilde{T}%
(g\mathbf{y}), 
\]%
for every $g\in SO(3)$ and  $\mathbf{x}, \mathbf{y} \in \mathbb{S}^2.$  This means that the mean $ \mathbf{E}T(\theta ,\varphi )=constant$
 and that the covariance function $ \mathbf{E}T(\theta
,\varphi )T(\theta ^{\prime },\varphi ^{\prime })$ depends only on the
angular distance $\Theta =\Theta _{PQ}$ between the points $P=(\theta
,\varphi )$ and $Q=(\theta ^{\prime },\varphi ^{\prime })$ on $\mathbb{S}^2.$ 
A wide class of non-isotropic random field models can be obtained by adding a deterministic component $T_{det}(\theta ,\varphi )$ to the field $T.$
 
As an example of a spherical random field we use the total column ozone data from the Nimbus-7 polar orbiting satellite, see more details in \cite{Cressie}. This data set provides measurements of the total amount of atmospheric ozone in a given
column of a $1^o$ latitude by $1.25^o$ longitude grid. The CSV file available from the website \url{https://hpc.niasra.uow.edu.au/ckan/dataset/tco} contains 173405 rows with the measurements recorded on the 1st of October, 1988.  We will be using the fields {\it lon: longitude, lat: latitude} and {\it ozone: TCO level 2 data}. 

First we demonstrate how to use \rcosmo to transform the geographic referenced ozone data to the HEALPix representation. The R code below loads the ozone data from the file {\it toms881001.csv} into R. Then geographic coordinates are transformed to spherical ones in radians. Finally, the function {\it HPDataFrame} creates an \rcosmo object at the resolution 2048, i.e. on the 50,331,648 nodes grid.

\begin{verbatim}
> library(rcosmo)
> library(celestial)
> totalozone <- read.csv("toms881001.csv")
> sph <- geo2sph(data.frame(lon = pi/180*totalozone$lon,  lat =
  pi/180*totalozone$lat))
> df1 <- data.frame(phi = sph$phi, theta = sph$theta,  
  I = totalozone$ozone)
> hp <- HPDataFrame(df1, auto.spix = TRUE, delete.duplicates 
  = TRUE, nside = 2048)
\end{verbatim}

Now we transform the result to an object of the CMBDataFrame class, which is the main class for statistical and geometric analysis in \rcosmo.
\begin{verbatim}
> cmb <- as.CMBDataFrame(hp)
\end{verbatim}

To visualise the data we first centre them by subtracting the mean and then rescale to use the \rcosmo colour scheme.
\begin{verbatim}
> cmb$I1 <- (cmb$I-mean(cmb$I))/100000
> plot(cmb, intensities = "I1", back.col = "white", size = 10)
\end{verbatim}

Now we add the coastline of Australia to the obtained 3d plot. We use the R package {\it map} to extract longitude and latitude coordinates of the Australian boarder. Then, similarly to the above code we transform the border coordinates to a CMBDataFrame object and plot it on the ozone map. 

\begin{verbatim}
> library(maps)
> library(mapdata)
> aus<-map("worldHires", "Australia", mar=c(0,0,0,0), plot =FALSE)
> aus1 <- data.frame(aus$x,aus$y)
> aus1 <- aus1[complete.cases(aus1),]
> sph1 <- geo2sph(data.frame(lon = pi/180*(aus1[,1]+180),
  lat = pi/180*(aus1[,2])))
> df2 <- data.frame(phi = sph1$phi,theta = sph1$theta, I = 1)
> hp1 <- HPDataFrame(df2, auto.spix = TRUE, delete.duplicates
  = TRUE, nside = 2048)
> cmb1 <- as.CMBDataFrame(hp1)
> plot(cmb1,  size = 10, col = "black", add=TRUE)
\end{verbatim}

\begin{figure}[h!]\vspace{-5mm}
\includegraphics[trim={7cm 4.5cm 5cm 5cm},clip, width=1.1\textwidth]{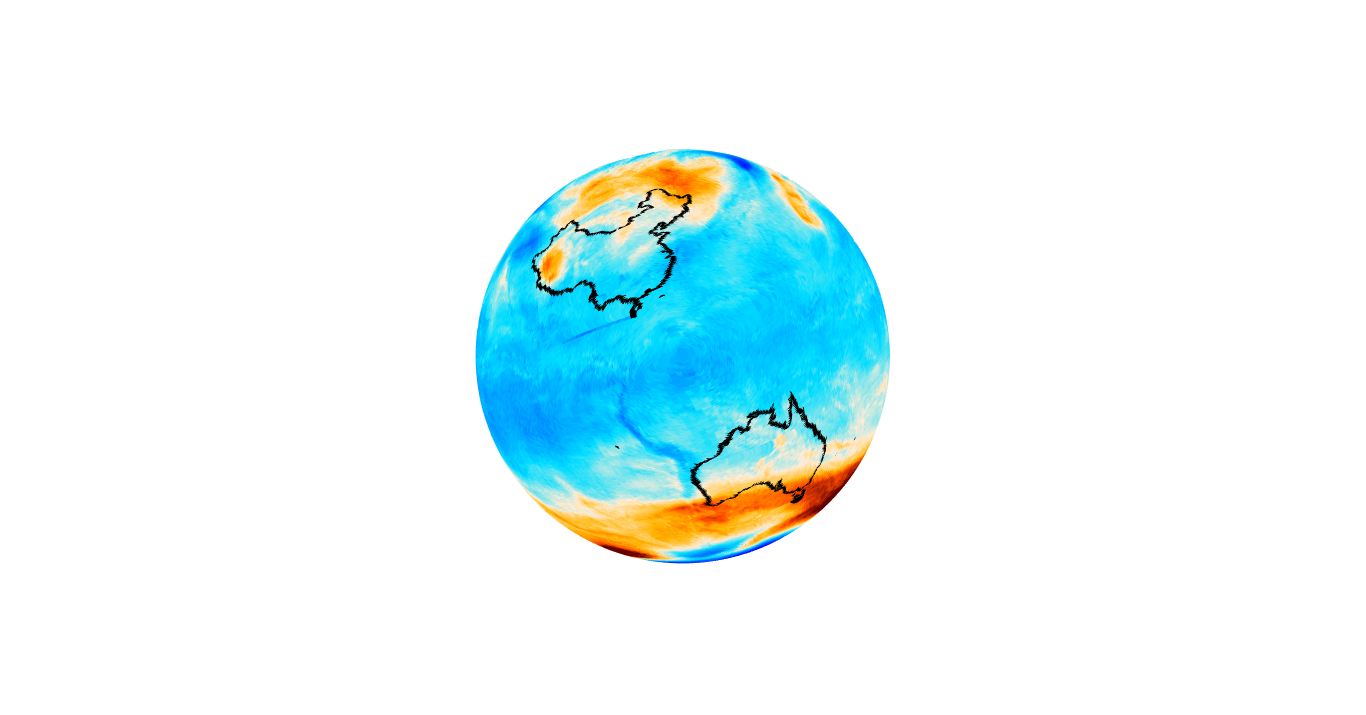}
\caption{Total column ozone map with Australia and China boundaries.} \label{ozone1}
\end{figure}

Setting the country to China 
\begin{verbatim}
> chi <- map("worldHires", "China", mar=c(0,0,0,0), plot =FALSE)
\end{verbatim}
and repeating the above commands will add the boundaries of China to the plot. The result is shown in Figure~\ref{ozone1}.\\

Now the data are in the HEALPix format and \rcosmo functions can be used to analyse them.  For example, the following code first computes the sample mean {\it alpha} of the total column ozone data. Then \rcosmo commands {\it exprob} and {\it extrCMB} estimate the exceedance probability above the level {\it alpha}  and get  three largest ozone values and their locations within the spherical window {\it win1}. 

\begin{verbatim}
> alpha <- mean(cmb[,"I", drop = TRUE])
> alpha
[1] 298.4333
> win1 <- CMBWindow(theta = c(0,pi/2,pi/2), phi = c(0,0,pi/2))
> exprob(cmb, win1, alpha,intensities = "I")
[1] 0.3557902
> extrCMB(cmb, win1, 3, intensities = "I")
A CMBDataFrame
# A tibble: 3 x 4
      I theta   phi       I1
  <dbl> <dbl> <dbl>    <dbl>
1  179.  3.07  3.32 -0.00119
2  180.  2.96  4.21 -0.00119
3  180.  2.94  4.27 -0.00118
\end{verbatim}

To compute the estimated entropy for the ozone measurements within the region {\it win1} one can use
\begin{verbatim}
> entropyCMB(cmb, win1)
[1] 2.214391
\end{verbatim}\vspace{-4mm}

\section{Point pattern data}\label{sec4}
In this section we demonstrate \rcosmo handling of geographic point pattern data. Comparing to continuous geographic data these points are not densely regularly spaced and often have random spatial locations. Some classical examples include studies of settlement distributions, locations of trees, seismological events, data aggregated over a set of zones to specific "central" locations. Spatial point data are also common in geographical epidemiology studies that deal with disease mapping,  clustering  and finding locations of possible sources. These data are usually represented and visualised as GIS vector images.

In statistical applications random spatial patterns of points are often modelled by spatial point processes. Points usually represent locations of  objects and the associated marks are used to record properties of these objects. 

A {\it spatial point process} $X$ is a random countable subset of the  sphere $\mathbb{S}^2.$ This process is a measurable mapping defined on the probability space $(\Omega ,\mathcal{F},\mathbf{P})$ and taking values in finite/countable sets of points from $\mathbb{S}^2.$ For every Borel subset $A \subset \mathbb{S}^2$ the corresponding random variable $N(A)$ denotes the number of points in this subset. For simplicity we restrict our consideration to simple point processes that have realizations with no coincident points.

The distribution of a point process $X$ is  defined by the joint distributions of the numbers of points $(N(A_1),...,N(A_k))$  in the subsets $A_1,...,A_k\subset \mathbb{S}^2,$ $k\in \mathbb{N}.$
 A point process on $\mathbb{S}^2$ is isotropic if its distribution is invariant under the group $SO(3).$ 
The  mean measure of a point process $X$ assigns to every subset $A \subset \mathbb{S}^2$ the expected number of points in this subset.

The most popular point processes in applications are Poisson and Cox point processes. More details on the theory and applications of spatial point processes can be found in \cite{Baddeley}, \cite{Diggle} and the references therein.

As an example we use the Integrated Global Radiosonde Archive (IGRA) measurements. They were collected from radiosondes and pilot balloons at over 2700 stations, \cite{IGRA}. Observations include locations of stations, temperature, pressure, wind direction and speed, etc. For the following analysis we used latitudes and longitudes  of stations and their elevations above sea level. The TXT file {\it igra2-station-list.txt} with IGRA stations data was downloaded from the website  \url{https://www1.ncdc.noaa.gov/pub/data/igra/} and saved as a CSV file.

First, the records with missing information and values "-9999" were removed. Then the longitude was recorded using the range $[0,360] .$
\begin{verbatim}
> x <- read.csv("igra2-station-list.csv", header=FALSE)
> x1 <- x[,c("V2","V3","V4")]
> x1 <- x1[complete.cases(x1),]
> x1 <- x1[x1$V3>-300,]
> x1$V3 <- x1$V3 + 180
\end{verbatim}

Similar to Section~\ref{sec3}, data were transformed to the CMBDataFrame class with the variable "I" denoting stations' elevation above sea level.
\begin{verbatim}
> sph <- geo2sph(data.frame(lon = pi/180*x1$V3, lat = 
  pi/180*x1$V2))
> df1 <- data.frame(phi = sph$phi, theta = sph$theta, I = x1$V4)
> cmb <- as.CMBDataFrame(hp)
\end{verbatim}

After a rescaling the locations of IGRA stations were plotted with darker colours corresponding to higher elevated stations. 
\begin{verbatim}   
> cmb$I1 <- (cmb$I-mean(cmb$I))/1000000
> plot(cmb,intensities = "I1", size = 3, back.col = "white",
  add=TRUE)
\end{verbatim}\nopagebreak

\begin{figure}
\includegraphics[trim={10cm 5cm 5cm 5.5cm},clip, width=1.2\textwidth]{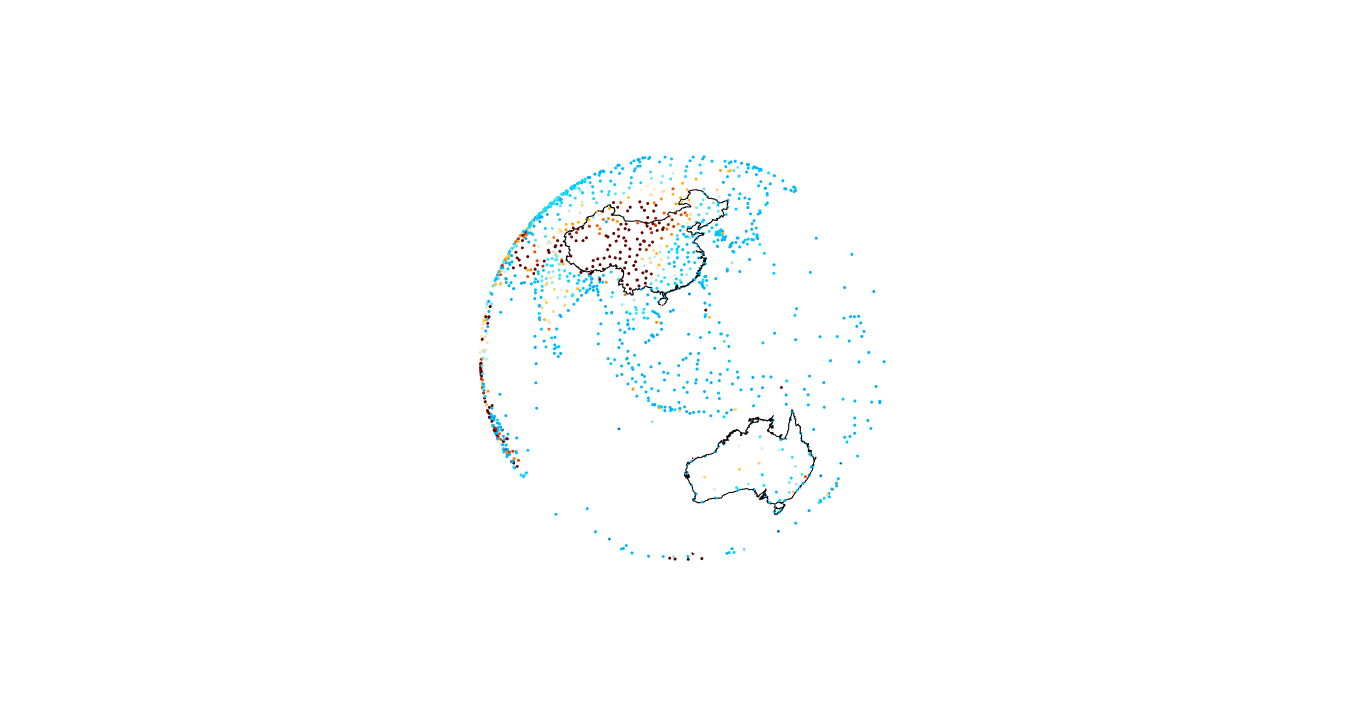}
\caption{Locations of IGRA stations and Australia and China boundaries.} \label{IGRA}
\end{figure}

Similar to Section~\ref{sec3} the boundaries of Australia and China were added to reference stations positions, see the resulting Figure~\ref{IGRA}. The figure suggests that the stations are not uniformly distributed over the globe and much higher elevated in China than in Australia.

As the data are in the HEALPix format a few \rcosmo functions were employed to analyse them. For example, the first Minkowski functional {\it fmf} can be used to estimate a relative area of HEALPix locations with the elevation above sea level. 
\begin{verbatim} 
> fmf(cmb, 0, intensities = "I")/(dim(cmb)[1]*pixelArea(cmb))
[1] 0.9869792
\end{verbatim}

The minimum angular geodesic distance between IGRA stations was computed by
\begin{verbatim}
> minDist(cmb)
[1] 0.00049973 
\end{verbatim}

\begin{figure}\vspace{-5mm}
\includegraphics[trim={0cm 0cm 0cm 0.5cm},clip, width=\textwidth]{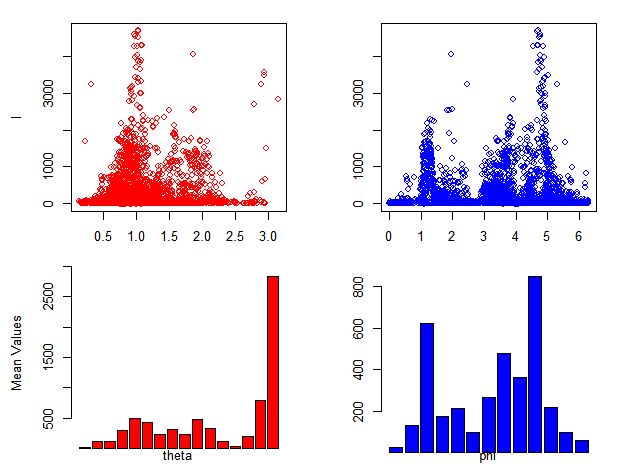}
\caption{Distribution of the elevation with respect to spherical angles.} \label{marg}
\end{figure}

The marginal distribution plots in Figure~\ref{marg} show that elevation departs from the uniform distribution with respect to geographic coordinates. 
\begin{verbatim}
> plotAngDis(cmb, intensities = "I")
\end{verbatim}
\vspace{-5mm}
\section{Directional data} 

This section demonstrates how to use \rcosmo with directional and shape data. In this publication we restrict our consideration only to star-shaped data. More details on other models  and methods in statistical directional and shape analysis can be found in~\cite{Ley} and \cite{Srivastava}. 

In contrast to  geographic data in Sections~\ref{sec3} and \ref{sec4}, directional data are not necessarily located on a sphere, but rather are observed in radial directions from a common centre. However,  they are usually indexed by points of the unit sphere.  Directional and shape data are common in various fields. For example, in geo-sciences (direction of the Earth’s magnetic pole, epicentres of earthquakes, directions of remnant rock magnetism), biology (movement directions of birds and fish,  animal orientation), in physics (optical axes of crystals, molecular links, sources of cosmic rays), etc.

The main statistical tools to model and investigate  directional data are {\it circular and spherical distributions and statistics,} see, for example, \cite{Ley}.  Probably the simplest spherical distribution is the uniform one with the constant density $1/(4\pi)$ for all $\mathbf{x}\in \mathbb{S}^2.$ This is the only distribution that is invariant under both rotations and reflections. An important directional statistic is the sample mean direction, which is computed as the direction of the sum $\sum_{i=1}^n\mathbf{x}_i$ of the observed set of unit vectors $\mathbf{x}_i\in \mathbb{S}^2,$ $i=1,...,n.$

Many of directional methods can be translated from spheres to {\it star-shaped surfaces,} with additional marks representing radial distances to observations. 
A~body $U$ in $\mathbb{R}^3$ is called star-shaped if there is a point $\mathbf{x}_0\in \mathbb{R}^3$ such that for every $\mathbf{x}\in U,$ the segment joining $\mathbf{x}_0$ and $\mathbf{x}$ belongs to $U.$ The corresponding body's surface is also called  star-shaped. The marks containing radial distances can be used to statistically investigate and compare various geometric properties of star-shaped data. For example, to estimate the mean directional asymmetry in a solid spherical angle $\omega$ one can use the excess from the overall mean distance  \[\frac{\sum_{x_i\in\omega}||\mathbf{x}_i -\mathbf{x}_0||}{\# i: x_i\in\omega}  \Big/\ \frac{\sum_{x_i\in U}  ||\mathbf{x}_i -\mathbf{x}_0||}{\# i: x_i\in U},\]
where $\#$ denotes a number of cases.

In this section we consider the shape data of the brain substructure amygdala studied in \cite{Chung}. Structural abnormalities of amygdala are related  to  functional impairment in autism. The data consist of amygdala MRI measurements  of 46~control and autistic persons and contain their group identifiers, age, left and right amygdala surface coordinates. The MATLAB file {\it  chung.2010.NI.mat} available from the website \url{http://pages.stat.wisc.edu/~mchung/research/amygdala/}  includes 2562 surface points for each person. 

First we load the full data set into R
\begin{verbatim} 
> library(R.matlab)
> mat <- R.matlab::readMat("chung.2010.NI.mat")
\end{verbatim}

\begin{figure}[th]\vspace{-4mm}
	\begin{minipage}{0.46\textwidth}
		\includegraphics[trim =  20cm 5cm 5cm 5cm, clip, width=1.5\textwidth]{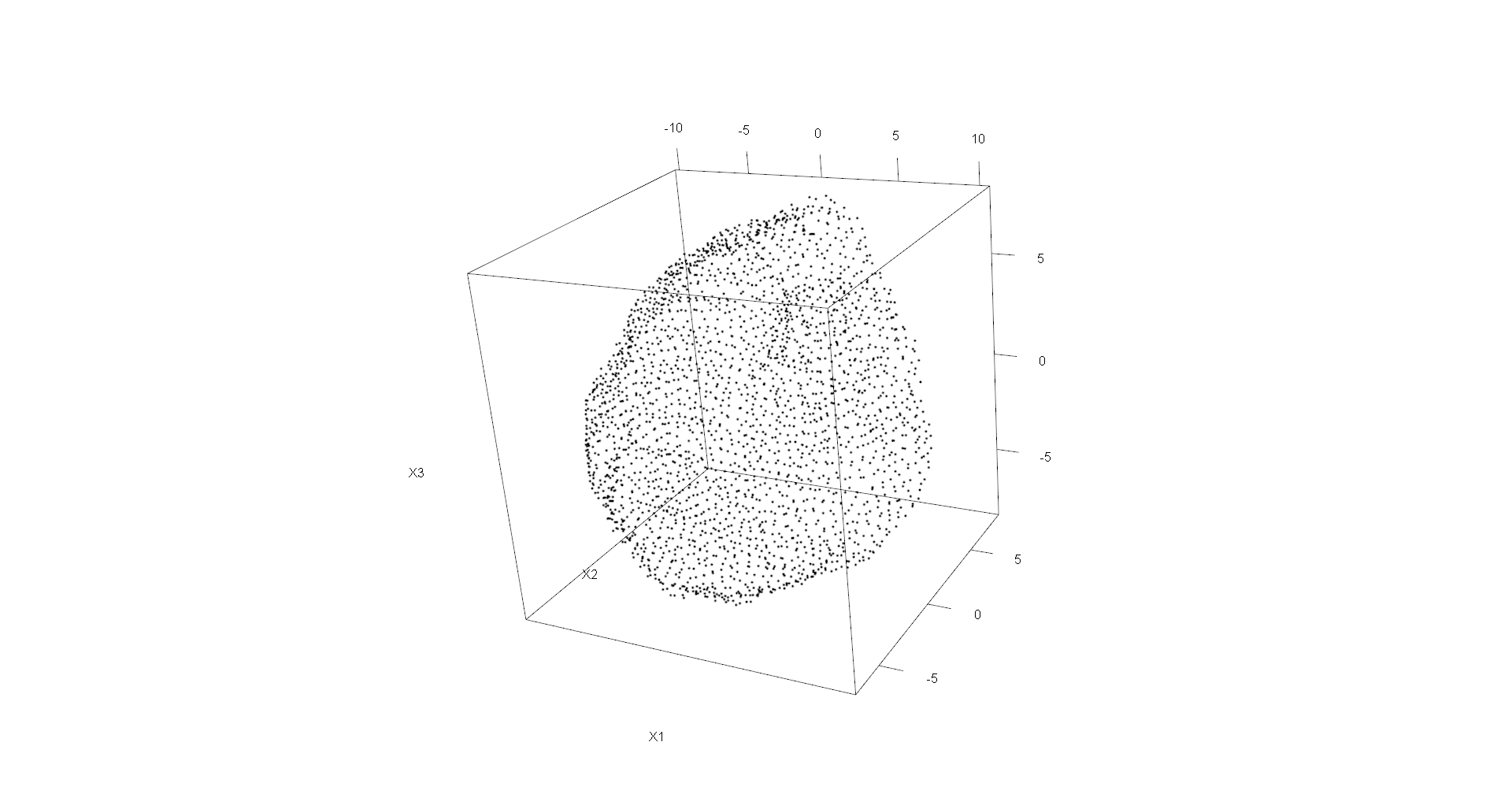}
	\end{minipage}\hspace{0.5cm}
	\begin{minipage}{0.46\textwidth}
		\includegraphics[trim =  20cm 5cm 5cm 5cm,clip, width=1.5\textwidth]{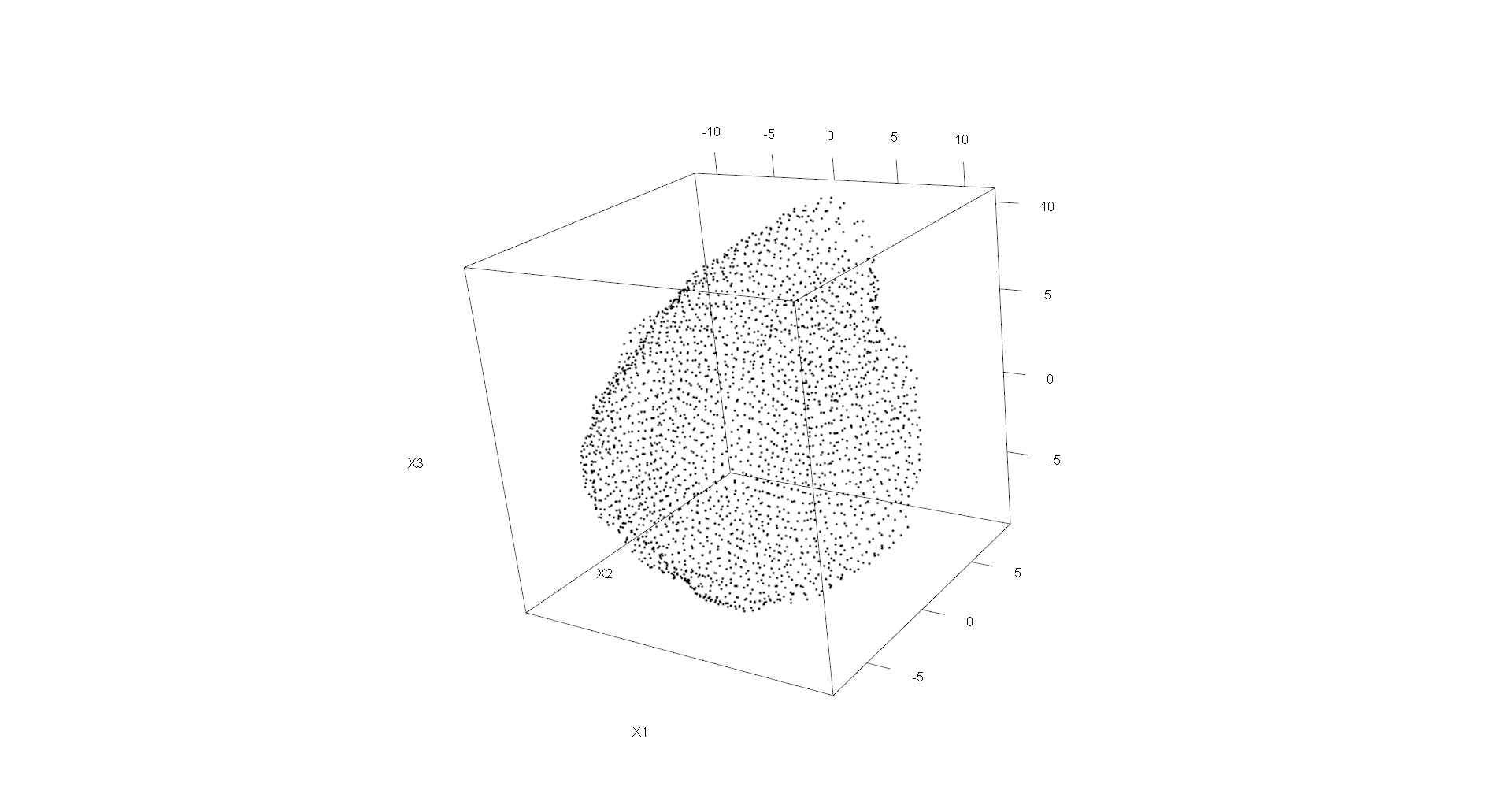}
	\end{minipage}
	\caption{Sampled points of left amygdala surfaces of persons 10 and 13.}\label{surf}
\end{figure}

Then we select two persons 10 and 13 (control and autistic) of the same age~17.  Cartesian coordinates of left amygdala sampled points of person 10   were transformed by first centring them and then converting to spherical coordinates. The corresponding 3d plot is shown in the first Figure~\ref{surf}.
\begin{verbatim} 
> p1 <- data.frame(mat$left.surf[10,,])
> p1 <- apply(p1, 2, function(y) y - mean(y))
> library(rgl)
> plot3d(p1)
> library(sphereplot)
> p1s <- as.data.frame(car2sph(p1, deg = FALSE))
> names(p1s) <- c("theta", "phi", "I")
\end{verbatim}

In contrast to Sections~\ref{sec3} and \ref{sec4}, now we let \rcosmo to find an {\it nside} resolution that separates points so that each belongs to a unique pixel. Then we save the data as a CMBDataFrame and create a new variable I1 with rescaled distances $||\mathbf{x}_i -\mathbf{x}_0||$ to use the \rcosmo colour scheme.
\begin{verbatim} 
> hp1 <- HPDataFrame(p1s, auto.spix = TRUE)
> cmb1 <- as.CMBDataFrame(hp1)
> cmb1
A CMBDataFrame
# A tibble: 2,562 x 3
I theta     phi
<dbl> <dbl>   <dbl>
1  7.82 0.470 0.0289 
2  8.24 0.474 3.05   
3  8.87 2.65  0.00577
4  8.15 2.67  3.10   
...
\end{verbatim}
\begin{verbatim} 
> pix(cmb1) <- pix(hp1)
> cmb1$I1 <- (cmb1$I-mean(cmb1$I))/1000
> plot(cmb1,intensities = "I1",back.col = "white", size = 3, 
  xlab = '', ylab = '', zlab = '')
\end{verbatim}

\begin{figure}[th]
	\begin{minipage}{0.49\textwidth}
		\includegraphics[trim = 0.5cm 20mm 10mm 3.5cm, clip, width=1\textwidth]{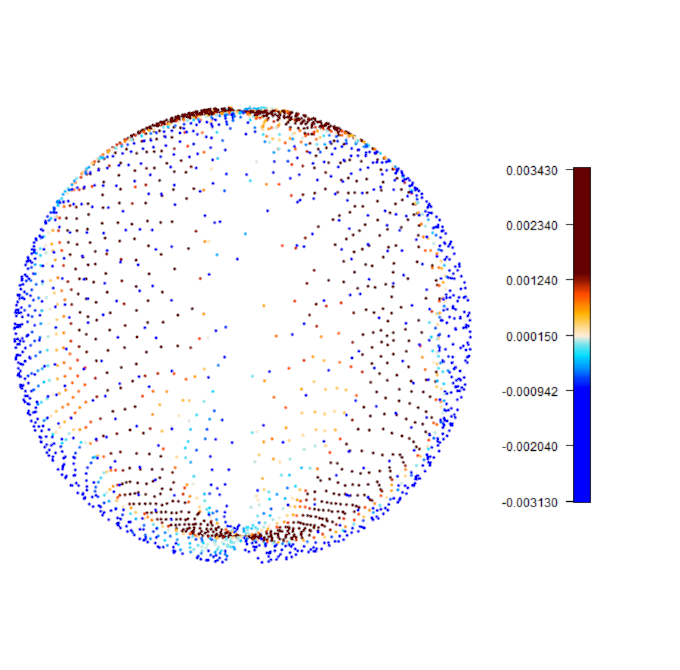}
	\end{minipage}\hspace{1mm}
	\begin{minipage}{0.49\textwidth}
		\includegraphics[trim = 0.5cm 20mm 15mm 3.5cm,clip, width=1\textwidth]{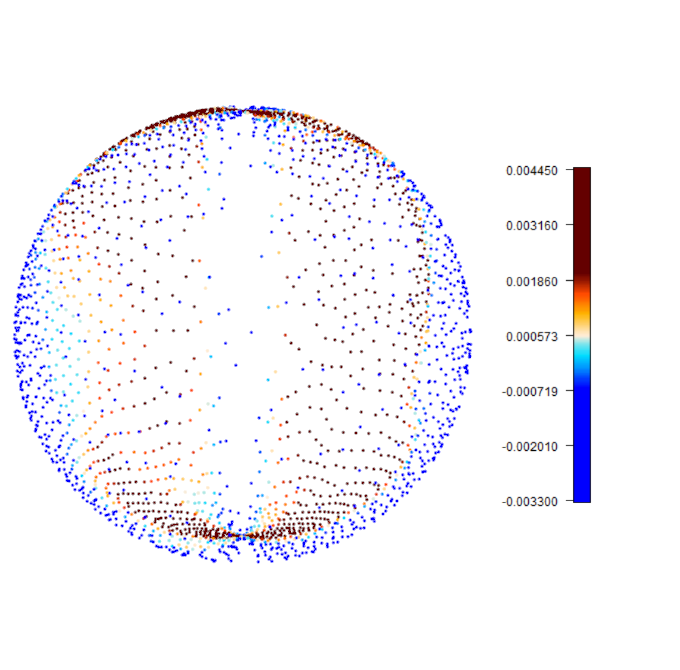}
	\end{minipage}
		\caption{Heat maps of $||\mathbf{x}_i -\mathbf{x}_0||$ for persons 10 and 13.} \label{splot}
\end{figure}

We repeat the same steps for the left amygdala of person 13.  The second Figure~\ref{surf} shows sampled points of the left amygdala of this person. The spherical plots in Figure~\ref{splot} use colours to represent the values of $||\mathbf{x}_i -\mathbf{x}_0||$ in the corresponding directions for each person.

\begin{figure}[tbh]\vspace{-4mm}
	\begin{minipage}{0.49\textwidth}
		\includegraphics[trim =  0.5cm 10mm 5mm 1.9cm, clip, width=1.0\textwidth, height=5cm]{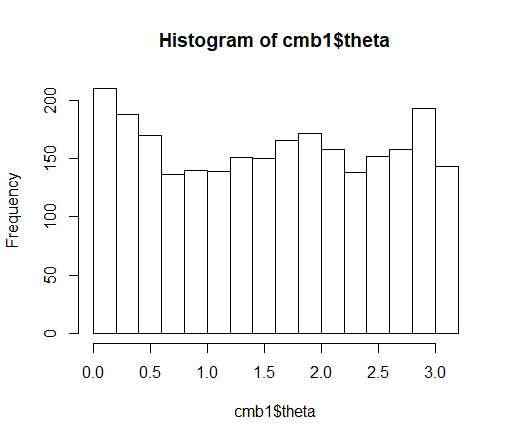}
	\end{minipage}\hspace{1mm}
	\begin{minipage}{0.49\textwidth}
		\includegraphics[trim = 0.5cm 10mm 5mm 1.9cm,clip, width=1\textwidth, height=5cm]{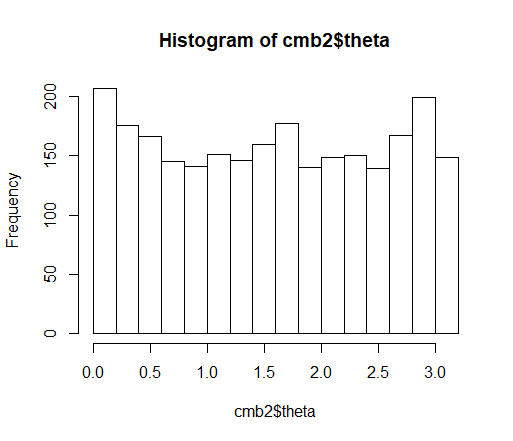}
	\end{minipage}
		\caption{Distributions of sampled points with respect to $\theta$ for persons 10 and 13.}\label{hist}
\end{figure}

To analyse and compare shapes of the amygdalae we first use directional histograms. For example, Figure~\ref{hist} suggests that directional distributions of sampled points with respect to $\theta$ may not significantly differ. Similar results were obtained for $\varphi$  directions. Thus, directional sampling rates of amygdalae for persons 10 and 13 look very similar.

However, basic statistical analysis of the variable I containing values of the sampled radial distances $||\mathbf{x}_i -\mathbf{x}_0||$ shows differences in the shapes of the control and autistic cases:
\setlength{\tabcolsep}{0.7em} 
{\renewcommand{\arraystretch}{1.5}
\begin{table}[h]\vspace{-3mm}
\begin{tabular}{|c|c|c|}
\hline
 Person   &  Mean &  First Minkowski functional   \\
 \hline
 10 &  7.525838 & 0.0003348093  \\
 \hline
 13 & 8.396525 &   0.0004266883  \\
 \hline
\end{tabular}
\caption {Basic statistics of the sampled radial distances for persons 10 and 13.}
\end{table}\vspace{-4mm}

Person 13 has larger amygdala. The area where the observed values of the distances  $||\mathbf{x}_i -\mathbf{x}_0||$ exceed the mean value \textit{mean(cmb1\$I)} for person 13 is larger by more than 27\%  than for person~10. 

To confirm that the difference between two subjects is not only in the amygdalae' sizes, but also in their shapes, one can study relative asymmetries. The \rcosmo command {\it CMBWindow} was used to select a spherical angle. Then, means of 10 largest values of $||\mathbf{x}_i -\mathbf{x}_0||$ in this angle were computed by the function {\it extrCMB} and normalised by the overall mean values.   
\begin{verbatim} 
> win1 <- CMBWindow(theta = c(pi/2,pi,pi/2), phi = c(0,0,pi/2))
> mean(extrCMB(cmb1, win1, 10)$I)/mean(cmb1$I)
[1] 0.6875167
> mean(extrCMB(cmb2, win1, 10)$I)/mean(cmb2$I)
[1] 0.75863
\end{verbatim}
The results suggest that not only absolute but also relative asymmetries of  amygdalae for the control and autistic persons 10 and 13 are different.

\section{Discussion}
This paper complements the detailed description of the structure and functionality of
the package \rcosmo in \cite{Fryer 2}. Here we only focus on potential applications of \rcosmo to non-CMB data recorded in non-HEALPix formats. 
  
It is demonstrated how radial or 3d spatial data in geographic or Cartesian coordinates can be transformed into the HEALPix format and analysed with the package \rcosmo. As one can use a very detailed HEALPix resolution (for example, the current CMB maps use Nside=2048 with more than 50 millions pixels) the impact of approximation errors of the coarseness of the HEALPix tessellation on statistical analyses can be made negligible. 

Applications are shown for the three different classes of non-HEALPix data that are often occur in practice: continuous geo-referenced fields, spatial points and star-shaped measurements. It would be important to prepare a similar guideline and analysis for another common spatial data type, spherical curves.

The presented results should serve as a demonstration and guideline of  bridging between data-types and formats  rather than detailed rigorous statistical analyses of the considered data. The readers can find further information on the scope of the package and implemented methods in \cite{Fryer 2}.

\section{Acknowledgements}
 We would like to thank V.V. Anh, P. Broadbridge, N. Leonenko, M. Li, I. Sloan, and Y. Wang for their discussions of CMB and spherical statistical methods, and J. Ryan for developing and extending the {\bf mmap} package. The authors are also grateful for the referees' comments, which helped to improve the style of the presentation.


\begin{thebibliography}{99}
	
\bibitem{Baddeley}  Baddeley, A.,  Rubak, E.,  Turner, R.:	Spatial Point Patterns.	Methodology and Applications with R. 
Chapman and Hall/CRC, New York (2015)

\bibitem{Chung} Chung, M.K., Worsley, K.J., Nacewicz, B.M., Dalton, K.M., Davidson, R.J.: General multivariate linear modeling of surface shapes using SurfStat. NeuroImage \textbf{53}, 491--505 (2010) \url{https://doi.org/10.1016/j.neuroimage.2010.06.032}

\bibitem{Cressie} Cressie, N., Johannesson, G.: Fixed rank kriging for very large spatial data sets. Journal of the Royal Statistical Society: Series B (Statistical Methodology) \textbf{70}(1), 209--226 (2008) \url{http://dx.doi.org/10.1111/j.1467-9868.2007.00633.x}

\bibitem{Diggle} Diggle, P.J.: Statistical Analysis of Spatial and Spatio-Temporal Point Patterns. Chapman and Hall/CRC, New York (2013)

	\bibitem{Fryer}  Fryer, D., Olenko, A., Li,  M.,  Wang, Yu.:
	 rcosmo: Cosmic Microwave Background Data
	Analysis. R package version 1.1.0.
	\url{https://CRAN.R-project.org/package=rcosmo} (2019)
	
\bibitem{Fryer 2} Fryer, D., Olenko, A., Li,  M.: rcosmo: R Package for Analysis of Spherical, HEALPix and Cosmological Data. submitted \url{https://arxiv.org/abs/1907.05648} (2019)

	
\bibitem{Gorski} Gorski, K.M., Hivon, E., Banday, A.J., Wandelt, B.D., Hansen, F.K., Reinecke, M., Bartelmann, M.: HEALPix: a framework for high-resolution discretization and fast analysis of data distributed on the sphere. The Astrophysical Journal, \textbf{622}(2), 759--771 (2005) \url{https://doi.org/10.1086/427976}
	
	\bibitem{healpix}
	HEALPix. Data Analysis, Simulations and Visualization on the Sphere. \url{https://healpix.sourceforge.io/}. Last accessed 30 May 2019
	
	\bibitem{healpix1}
	Healpy documentation homepage. \url{https://healpy.readthedocs.io/}. Last accessed 30 May 2019
		
\bibitem{healpix2}
	HEALPix Library for MATLAB. \url{http://sufoo.c.ooco.jp/program/healpix.html}. Last accessed 30 May 2019
	
\bibitem{IGRA}
	Integrated Global Radiosonde Archive homepage. \url{https://www.ncdc.noaa.gov/data-access/weather-balloon/integrated-global-radiosonde-archive}. Last accessed 30 May 2019
		
	\bibitem{Ivanov}  Ivanov, A. V.,  Leonenko, N. N.:  {Statistical Analysis
		of Random Fields}. Kluwer Academic Publishers, Dordrecht (1989)
		
	\bibitem{Ley} Ley, C., Verdebout, T.: Modern Directional Statistics. CRC Press, Boca Raton, FL (2017)
		
	\bibitem{MarinucciPeccati11} Marinucci, D.,  Peccati, G.:  {Random
		Fields on the Sphere. Representation, Limit Theorems and Cosmological
		Applications}. Cambridge University Press, Cambridge (2011) 
		
		\bibitem{Srivastava} Srivastava, A., Klassen, E.P.: Functional and Shape Data Analysis. Springer, New York (2016)
	
	\bibitem{Yadrenko}  Yadrenko, M. I.:  {Spectral Theory of Random Fields}.
	Optimization Software Inc., New York (1983)
\end{thebibliography}
\end{document}